\documentstyle[aps]{revtex} % for preprint form, small characters
\begin{document}
%\twocolumn
\draft
\title{Electronic Raman scattering in $HgBa_{2}Ca_{2}Cu_{3}O_{8+\delta } $
single crystals. Analysis of the superconducting state}
\author{A. Sacuto, R. Combescot \dag, N. Bontemps and C. A. M\"uller}
\address{Laboratoire de Physique de la Mati\`ere Condens\'ee,
\dag Laboratoire de Physique Statistique, Ecole Normale Sup\'erieure,
24 rue Lhomond, 75005 Paris , France}
\author{V. Viallet and D. Colson}
\address{Physique de l'Etat Condens\'e, DRECAM/SPEC, CEA, Saclay,
91191 Gif sur Yvette , France}
\date{Received \today}
\maketitle

\begin{abstract}
Electronic Raman scattering measurements have been performed 
on $HgBa_{2}Ca_{2}Cu_{3}O_{8+\delta } $  single crystals in the 
superconducting state. Pure electronic Raman spectra with no 
phonon structures hindering the analysis of the electronic continuum 
have been obtained. As a consequence, the spectra in the pure 
$B_{1g } $ and $B_{2g } $ symmetries are directly and reliably analyzed  
and the pure $A_{1g } $ contribution can be easily identified. 
Below the critical temperature $T_{c}, $ two electronic structures 
at $2\Delta \sim 6.4 k_{B}T_{c} $ and $2\Delta \sim 9.4 $
$k_{B}T_{c} $ are clearly seen. 
Both are observed simultaneously in pure $A_{1g} $ symmetry, the 
highest energy one being located at the energy of the $B_{1g} $ 
maximum. These two maxima disappear at $T_{c} $ and do not soften 
significantly as the temperature is raised up to $T_{c}. $ The 
low energy frequency dependence of the $B_{1g } $  electronic 
response is strongly linear, for various excitation lines in 
the 476.5 to 647.1 nm range. Such experimental data cannot be reconciled 
with a pure $d_{x^{2}-y^{2}} $  symmetry. Instead, they strongly
advocate in favor of an anisotropic superconducting gap with two distinct 
gap maxima and of nodes existing outside the [110] and 
[1,$\bar{1}$,0] directions in {\bf k}-space. We discuss in detail the 
simplest order parameter compatible with our experimental findings.
\end{abstract}

%\begin{multicols}{2}
%\narrowtext

\section{INTRODUCTION}
Since the last few years, identifying the symmetry of the pairing 
state has been the expected major step towards an understanding 
of high $ T_{c}$  superconductivity. The controversy is not 
yet resolved: while there seems to be an agreement on the existence 
of nodes in the gap, some experiments appear to advocate in 
favor of a pure $d_{x^{2}-y^{2}} $ symmetry \cite{[1]}, whereas others 
seem to exhibit a significant s-wave contribution \cite{[2]}. With 
respect to this problem, inelastic light scattering has been 
shown very early to be a powerful tool because, besides probing 
the bulk (in contrast with photoemission and tunneling), the 
polarization selection rules on the incident and scattered light 
make the spectra sensitive to the wave-vector of the electronic 
excitations \cite{[3],[4],[5],[6]}. A theoretical approach of electronic
Raman scattering has been developed, in the case of $d_{x^{2}-y^{2}} $ 
symmetry. The emphasis is put on the low energy spectrum, which 
is expected in this case to exhibit well defined power laws, 
according to the symmetry channel, e.g. the polarization of 
the incoming and outgoing light. More precisely, the so-called 
$A_{1g} $ and $B_{2g} $ symmetry should display a linear dependence 
versus frequency, as these channels are sensitive to the nodes 
in the $k_{x} = k_{y} $ direction, whereas the $B_{1g} $ symmetry probes 
mainly directions around the $k_{x}=0, $ $k_{y}=0 $ directions 
and should display a $\omega ^{3} $ 
frequency dependence \cite{[6]}. Such calculations omit other excitations 
such as phonons. The experimental major difficulty lies then 
in the fact that in most investigated compounds, e.g.
$La_{2-x }Sr_{x }CuO_{4}, $  
$YBa_{2}Cu_{3}O_{7-\delta } $ (Y-123), $Bi_{2}Sr_{2}CaCu_{2}O_{8+\delta } $ 
 (Bi-2212) and $Tl_{2}Ba_{2}CuO_{6+\delta } $ (Tl-2201), a large phonon 
contribution is superimposed to the electronic excitations in 
particular at low frequency and has therefore to be subtracted, 
which precludes an accurate determination of the shape of the 
electronic contribution in the superconducting state
\cite{[7],[8],[9],[10],[11],[12],[13],[14],[15]}. 
Despite this difficulty, in a number of reports, a satisfactory 
agreement is claimed to be found between the electronic response 
obtained after subtraction of the phonon lines, and the $d_{x^{2}-y^{2}} $
model \cite{[6],[12],[14]}

In this work, we report \emph{pure} electronic Raman spectra in
$HgBa_{2}Ca_{2}Cu_{3}O_{8+\delta } $
(Hg-1223) single crystals, belonging
to the highest $ T_{c}$ cuprate family. This provides a large spectral range
below the superconducting gap frequency where the low energy electronic
spectrum can be analyzed. Moreover, within this range, no subtraction of
the phonons is required to discuss the electronic response \cite{[16]}. Indeed,
as shown in our previous work [16-18], the very low intensity of the Raman
phonon peaks for electric fields within the $CuO_{2}$ planes allows a reliable
analysis of the electronic excitations in the superconducting state. The
crystallographic structure of Hg-1223 is tetragonal ($^{1}D_{4h}$) \cite{[19]}, which
yields an unambiguous comparison with theoretical calculations based on
tetragonal symmetry, without the complications due to the orthorhombic
distortion as in Y-123 \cite{[15]}.
Hg-1223 is therefore particularly well suited
for the study of the electronic Raman scattering. Our most striking
results are i) the existence of two electronic maxima at
$2\Delta = 6.4 k_{B}T_{c} $ and $2\Delta = 9.4 k_{B}T_{c} $,
both detected in $A_{1g}$ symmetry. These two maxima disappear
at $ T_{c}$. ii) the low frequency behavior of the $B_{1g}$ spectrum
displays what we believe to be an intrinsic linear term, which is observed
for all excitation energies used in this work (476.5-647.1 nm spectral
range). Such observations imply that the gap is strongly anisotropic with
two distinct maxima and that nodes exist outside the [110] and
[1,$\bar{1}$,0] directions. We argue that our results, which are at odds
with a pure $d_{x^{2}-y^{2}} $ gap symmetry, imply an order parameter
with two maximum gap values and eight nodes.

\section{Electronic Raman scattering in the framework of BCS theory}
We briefly recall the standard theoretical framework for electronic
Raman scattering which we will use in the analysis of our experimental
results \cite{[3]}. Electronic Raman scattering is a process of inelastic
light scattering where an incident photon is absorbed by the crystal
and a scattered one is emitted, with the simultaneous creation (Stokes)
or annihilation (anti-Stokes) of an electronic excitation. Here we will
only have to deal with Stokes processes. Since two photons are coming
into play, Raman scattering is a second order process in the
electromagnetic field. This second order effective interaction with
electronic excitations comes both from a direct second order term in
the interaction Hamiltonian, and from a first order term treated up
to second order in perturbation. As introduced by Abrikosov and
Genkin \cite{[4]}, one may consider that both terms can be gathered in a
single effective second order term in the Hamiltonian which can be
written as:
\begin{eqnarray}
{H}_{R}={{e}^{2} \over m}\left\langle{{A}_{S}{A}_{L}}
\right\rangle{e}^{-i\omega t}{\hat{\rho }}_{q}
\label{eq1}
\end{eqnarray}
where  e  is the electronic charge and  m  its bare mass. $ A_{L}$
and $ A_{S}$ are the vector potentials of the incoming laser and
scattered light, and the bracket
is for the proper matrix element over the photons states. The
difference between the incident and the scattered photon frequencies
is noted $\omega = \omega_{L} -\omega_{S} $ , and the difference between
the photon momenta is $ {\bf q} = {\bf k}_{L} -{\bf k}_{S} $. The
operator ${\hat{\rho }}_{q}$ is given by :
\begin{eqnarray}
{\hat{\rho }}_{q}=\sum\nolimits\limits_{{n}_{f},{n}_{i},\bf k}^{}
{\rm \gamma }_{{n}_{f},{n}_{i},\bf k}
{\rm c}_{{n}_{f},\bf k\rm +\bf q}^{\rm +}{c}_{{n}_{i},\bf k}
\label{eq2}
\end{eqnarray}
It is quite similar to the standard density operator, the only difference
being the ${\gamma }_{{n}_{f},{n}_{i},\bf k}$ for the scattering process,
known as the Raman vertex, which is given explicitely by :
\begin{eqnarray}
\lefteqn{{\gamma }_{{n}_{f},{n}_{i},\bf k}\rm
={\bf e}_{\rm S}.{\bf e}_{\rm L}{\bf
\delta }_{{\rm n}_{f},{n}_{i}}}  \nonumber \\
\nonumber  \\ [.1cm]
 & & +{1 \over \hbar m}\sum\nolimits\limits_{{n}_{m}}^{}
{\left\langle{{\rm n}_{f},\bf k\rm +\bf q\rm
 \left|{{e}^{-i{\bf k}_{\rm s}.\bf r}\rm {\bf e}_{\rm
S}.\bf p}\right|{\rm n}_{m},\bf k\rm +{\bf
k}_{L}}\right\rangle\left\langle{{\rm n}_{m},\bf k\rm
+{\bf k}_{L}\rm \left|{{e}^{i{\bf k}_{\rm L}.\bf
r}\rm {\bf e}_{\rm L}.\bf p}\right|{\rm n}_{i},\bf
k}\right\rangle \over {\rm \varepsilon }_{{n}_{i},\bf
k}\rm -{\varepsilon }_{{n}_{m},\bf k\rm +{\bf k}_{L}}\rm
+{\omega }_{L}+i\eta }+(L\leftrightarrow S)
\label{eq3}
\end{eqnarray}
It is quite important that this Raman vertex has a $\bf k$-dependence,
otherwise electronic Raman scattering would essentially vanish, for
symmetry reasons in $B_{1g}$ and $B_{2g}$ symmetry, and because of screening
in $A_{1g}$ symmetry.

The interest of the reduced form in Eq.(2) is that we have to deal only
with electrons in a single band. In particular when the photon energies
$\omega_{L}$ and $\omega_{S}$  are neglected in the denominator of Eq.(3),
it can be shown that the Raman vertex is identical to the contraction of the
${\bf k}$ -dependent inverse effective mass tensor of this band with the
polarization vectors of the incident and the scattered light. This
latter reduced formula is sometimes used to make explicit calculations
of the Raman vertex from an assumed knowledge of the band structure.
In general a single band is assumed at the Fermi level. It is not obvious
at all that this is reasonable in high $T_{c}$ compounds, although one
may argue that the Raman process is dominated by a single band
(this is strongly suggested by the similarity of electronic
Raman scattering between various high-$T_{c}$ superconductors).

Moreover the effective mass formula assumes that the energy of all
the states belonging to other bands are far enough from the Fermi
level in order to neglect the photon energies in the denominator.
While this approximation is reasonable for simple metals (one band
is near the Fermi level and the other bands are far away), it appears
much more doubtful in high-$T_{c}$ superconductors. In particular it is known
that the absorption is important at the incident frequency, linked
to interband transitions. This invalidates the approximation of
neglecting the photon energies and implies instead the existence of
resonant terms. This is actually coherent with what we observe
experimentally, since the electronic Raman scattering varies in an
important way as a function of the incident photon energy (as we will
see in section V.2). Moreover, in order to deal consistently with
these quasi-resonant terms, we have to take into account the finite
lifetime of the highly excited intermediate states. We expect, on the
basis of optical data, that this relaxation rate is large, of order of
a few tenths of eV. All this makes it clear that the effective mass
approximation is not valid in our experimental situation. Therefore
we will rather consider that the Raman vertex is actually an
effective quantity which depends not only on the wavevector $\bf{k}$,
but also on the incident and scattered photon energies. For simplicity
we will not indicate explicitly this dependence in the following, but
we will consider it in the analysis of our data.

The differential cross section is then obtained in terms of the
generalized susceptibility  $\chi (\beta,\omega,{\bf q} )$
corresponding to the generalized density Eq.(2).
It is related to the imaginary part $\chi " (\beta,\omega,{\bf q}) $
of this susceptibility by :
\begin{eqnarray}
{{d}^{2}\sigma  \over d\Omega d\omega }={{r}_{0}^{2}
\over \pi (1-{e}^{-\beta \omega })}{{\omega }_{S}
\over {\omega }_{L}}\chi "(\beta ,\omega ,\bf q\rm )
\label{eq4}
\end{eqnarray}
where $r_{0} = e^{2}/ 4 \pi \epsilon_{0} m c^{2}$ .
The standard procedure is then to calculate
$\chi (\beta,\omega,{\bf q})$ in the
superconducting state within the single bubble approximation.
Moreover it is appropriate to neglect the momentum transfer $\bf{q}$
because it is small compared to the inverse coherence length and
almost perpendicular to the $CuO_{2}$ planes. One finds then within BCS
theory and in the limit of low temperature $ T \rightarrow 0 $ ,
which is relevant for our experiments \cite{[3]} :
\begin{eqnarray}
\chi "(\omega )={2\pi {N}_{F} \over \omega }
Re\left\langle{{{\left|{{\gamma }_{\bf
k}}\right|}^{\rm 2}{\Delta }_{\bf k}^{\rm 2}
\over {({\omega }^{2}-4{\Delta }_{\bf k}^{\rm 2})}^{1/2}}}\right\rangle
\label{eq5}
\end{eqnarray}
$N_{F} $ is the density of states for both spin orientations at 
the Fermi level, and the brackets indicate an  average over 
the Fermi surface. $\Delta_{\bf k}$ 
stands for the superconducting, $\bf{k}$-dependent gap. For our purpose 
Eq.(5) has two related major drawbacks. First it leads to a 
vanishing Raman scattering in the limit of large frequency, 
where we should recover the physical properties of the normal 
state. It also leads in general to a singular behavior in the 
vicinity of the thresholds for creation of a pair of excitations, 
which occurs at twice the maxima of the gap $\Delta_{\bf k}$ .
This is in contrast with our experimental results. As we 
will see, we obtain smooth spectral shapes, and, in the high frequency 
regime, we recover a strong Raman scattering, which is well known 
in the normal state. These two discrepancies between experiment 
and theory can be ascribed to the fact that no collisional damping 
of any kind is included in the process of deriving Eq.(5). Nevertheless 
this is not so important here since our purpose is much more 
to make a qualitative analysis of our results than to provide 
a full quantitative account  of experiment. However it is clear 
that inclusion of lifetime effects in  the theory is quite desirable 
to provide a more accurate account of experimental results. 
Work in this direction is in progress and will be reported elsewhere.

Finally there is a physical effect, essential for a proper 
analysis of Raman scattering, which we have not included up 
to now. Indeed if we had a constant Raman vertex, we would merely 
look for a density-density response. As pointed out already 
by Wolff \cite{ecran,[4]} , perfect screening in metals at long wavelength 
implies that in this case Raman scattering would be essentially 
zero. Therefore it is necessary to include screening by the 
Coulomb interaction in order to obtain a proper physical result. 
In the limit of perfect screening, the result is : 
\begin{eqnarray}
{\chi }^{"}(\omega )=Re\left[{{\chi }_{\gamma
\gamma }(\omega )-{{\chi }_{\gamma 1}(\omega ){\chi }_{1\gamma }
(\omega ) \over {\chi }_{11}(\omega )}}\right]
\label{eq6}
\end{eqnarray}
where ${\chi }_{\gamma \delta }$ is defined by :
\begin{eqnarray}
{\chi }_{\gamma \delta }(\omega )={2\pi {N}_{F}
\over \omega }\left\langle{{{\gamma }_{\bf k}{\rm
\delta }_{\bf k}^{\star }{\rm \Delta }_{\bf k}^{\rm 2}
\over {({\omega }^{2}-4{\Delta }_{\bf k}^{\rm 2})}^{1/2}}}\right\rangle
\label{eq7}
\end{eqnarray}
Here ${\gamma }_{\bf k}$ and ${\delta }_{\bf k}$
are general vertices, 1 being the 
proper vertex for a true density interaction ( this comes in 
Eq.(6) because only this true density is screened by Coulomb 
interaction ). In $B_{1g} $ and $B_{2g} $ symmetry, $\chi _{\gamma 1}$ is 
zero for symmetry reasons, the last term in Eq.(6) drops out 
and we merely recover Eq.(5). In other words, the $B_{1g} $ and 
$B_{2g} $ symmetry are not screened because the average effective 
density is zero in these cases. On the other hand this is not 
so for the $A_{1g} $ symmetry, and screening has to be properly 
taken into account by making use of Eq.(6). Nevertheless one 
sees easily from this equation that any {\bf k}-independent 
contribution to the Raman vertex drops out of the result. Physically 
such a contribution is irrelevant because it is fully screened 
out. 

\section{CRYSTAL CHARACTERIZATION}
The measurements were carried on two single crystals labelled 
A and B. These crystals were grown by a single step synthesis 
as previously described for Hg-1223 \cite{[20]}. They are parallelepipeds 
with typical 0.5 x 0.5 $mm^{2} $  cross section and thickness 
0.3 mm. The [100] crystallographic direction lies at $45^{0}$ of the 
edge of the square and the [001] direction is normal to the 
surface. They were characterized by X-ray diffraction and wave 
length dispersive spectrometry \cite{[21]}. Identification of the proper 
phase requires several experimental characterization techniques, 
and one may be easily mislead in particular by the critical 
temperature. The presence of intergrowth phases, corresponding 
to different stacking numbers of the $CuO_{2} $ planes along the 
{\it c}-axis has been established \cite{[21]}. This was detected through: 
i) the presence of diffusion streaks in the oscillation photographs, 
ii) the lattice parameter values obtained from an Enraf-Nonius 
CAD-4 diffractometer at 293 K , iii) the chemical analysis carried 
out using a Camebax SX50 electron probe. The electron beam probes 
1 $\mu m^{3} $ of material at each measurement (the probed depth 
 is approximatively 10 times the penetration 
depth of the excitation light). 20 measurements were performed 
for the A crystal and 40 for the B crystal. The A crystal exhibits 
an overall composition of $Hg_{0.88}Ba_{2}Ca_{1.92}Cu_{2.88}O_{8+\delta }. $ 
The under-stoechiometry in Ca and Cu can be assigned to the 
existence of Hg-1212 intergrowth phase. The B crystal has the 
following composition: $Hg_{0.96}Ba_{2}Ca_{2.26}Cu_{3.5}O_{8+\delta }, $ 
which reveals a mixture of Hg-1223 and Hg-1234. A detailed X-ray 
analysis has been performed on the A crystal. The lattice parameters 
deduced from the diffractometer are a=3.844(2)$\rm \AA$  and
c=15.72(3)$\rm \AA$, 
hence the c parameter of this crystal is slightly smaller than 
the perfectly stoechiometric Hg-1223 (c=15.831 \AA) and definitely 
much larger than c=12.71$\rm \AA $  for Hg-1212. Therefore, according 
to the composition and the lattice parameters and in agreement 
with the chemical probe, we may consider that we are dealing essentially
with the Hg-1223 compound in both A and B crystals.

The crystals are slightly under-doped, with $\delta  $ less than 0.1. 
$\delta  $ refers to an excess of oxygen atoms randomly arranged within 
the Hg planes. Under-doping is further confirmed by DC magnetization 
measurements displayed in Fig.1. Although there is an onset 
of diamagnetism around 133 K, it appears that the main transition 
occurs at $T_{c} \sim 126 K. $ Optimal doping increases $\delta , $
hence introduces disorder in the Hg basal plane. Therefore an under-doped
system is closer to four-fold symmetry than an optimally doped one 
and is even more appropriate to symmetry considerations.

Finally we wish to point out that the extensive study of the 
crystal structure described above had not been completed for 
all the samples (coming from a single batch) by the time of 
our earlier work \cite{[16]}. One crystal was clearly identified as Hg-1212, 
and we eventually published the better quality spectra from 
another one, unaware of the fact, which is established by now, 
that two crystals coming from the same batch may be either Hg-1212 
or Hg-1223. The latter sample was subsequently analyzed and 
turned out to be, as explained above, Hg-1223. It is important 
to quote that actually the Raman spectra of the two samples 
studied in \cite{[16]} were similar. This will be shown in the experimental 
results.

\section{Experimental procedure}
Raman measurements were performed with a double monochromator 
using a single channel detection and the $Ar^{+} $  and $Kr^{+} $ 
laser lines. The spectral resolution was set at 3 $cm^{-1}. $ 
The crystals were mounted in vacuum $(10^{-5} $ mbar) on the 
cold finger of a liquid helium flow cryostat. The temperature 
was controlled by a Si diode located inside the cold finger. 
The incident laser spot is less than 100 $\mu m $ in diameter and 
the intensity onto the crystal surface was kept below $50 W.cm^{-2} $ 
 in order to avoid heating. At our lowest temperature, the difference 
in temperature between the cooled face and the illuminated face 
of the crystals was estimated to be $\sim $ 1 K , resulting in a temperature 
of 13 K inside the scattering volume, instead of a nominal 12 K 
temperature. The incidence angle was $60^{0}$. The scattered light 
was collected along the normal to the crystal surface. The polarization 
is denoted in the usual way: x[100] (a axis), y[010], z[001] 
(c axis), x'[110], y'[1$\bar{1}$0].
In order to compare our experimental data with the theoretical 
calculations [6,22], the pure $B_{2g} $ (xy) and $B_{1g} $ (x'y') 
symmetries are convenient and well suited for the investigation 
of a $d_{x^{2}-y^{2}} $  order parameter. To make this point clear, 
we recall briefly the symmetry properties: in the $B_{1g}  $ case,
the Raman vertex $\gamma_{\bf k}$ is zero by symmetry for $k_{x} = k_{y}$,
 so the electronic scattering is insensitive to the gap structure 
around $45^{0}$; in contrast, the  $k_{x} = 0 $ and  $k_{y} = 0 $
regions do contribute, giving weight to the gap $\Delta_{0}$
in these directions. Conversely, in the $B_{2g } $  case, the
Raman vertex $\gamma_{\bf k}$
is zero by symmetry for $k_{x} = 0 $ or $k_{y} = 0 $
or and non zero elsewhere, hence provides weight in the $k_{x} = k_{y}$
direction.

In our experimental procedure, we start by measuring the $B_{1g } $ 
channel. The incident angle being not zero, the crystal is then 
rotated by $45^{0}$ in order to get the $B_{2g } $ one, the polarizer 
and the analyzer remaining unchanged (see Fig. 2-a and 2-b). 
The $A_{1g }+B_{1g } $ and $A_{1g }+B_{2g } $ symmetries are
obtained 
from the (xx) and (x'x') polarizations respectively. In order 
to probe the same area for all symmetries, the impact of the 
laser beam onto the crystal surface was precisely located before 
turning the crystal. Finally, a very weak diffusive spot was 
carefully selected on each crystal in order to minimize the 
amount of spurious elastic scattering. 

Another possibility which guarantees that the location of the laser spot is 
unchanged for $B_{1g } $ and $B_{2g } $ symmetries, is to rotate 
the polarization instead of the crystal (see Fig. 2-c), when 
starting from pure $B_{1g} $ symmetry. After a $45^{0}$ rotation, the 
incident electric field no longer lies within the xy plane, 
which induces symmetry mixing for $B_{2g }. $ We have estimated 
the amount of mixing with $A_{1g }+B_{1g } $ (xx) and $E_{g }$ (xz),   
taking into account the high refractive index $(\sim 2), $ which brings 
the propagation of the beam inside the crystal closer to normal 
incidence; we found $\sim 5\% $ admixture of $A_{1g }+B_{1g } $ (xx) 
and $\sim 16\% $ of $E_{g }$ (xz).  We shall denote this mixed symmetry 
as $B'_{2g}$. Similarly, we can measure the $A'_{1g}$  
 channel (bringing the analyzer parallel to the polarizer), 
which comprises besides the $A_{1g }+B_{1g } $ spectrum, $\sim 5\% $ 
of $B_{2g } $ (xy) and $\sim 16\% $ of $E_{g }$ (xz). 

To get a complete set of reliable data, we have performed Raman 
measurements in both configurations: i) rotation of the crystal 
without changing the polarization, so as to get pure $B_{1g } $ 
and $B_{2g } $ symmetries and ii) rotation of the polarization 
without moving the crystal, in order to work with the same spot. 
Various spots have been probed on each crystal. All the resulting 
data are consistent one with the other.

\section{Experimental results}
The Hg-1223 raw spectra  at 13 K, in the three symmetries obtained 
with the 514.52 nm laser line by rotating the crystal are displayed 
in Fig. 3-a. These pure $A_{1g}+B_{1g}, $ $B_{1g} $ and $B_{2g} $ spectra 
were previously reported in ref \cite{[16]}. Figure 3-b shows the spectra 
obtained by rotating the polarization, yielding the $A'_{1g}$ 
and $B'_{2g}$ mixed symmetries. The two procedures (performed 
in the same A crystal) yield similar results for each symmetry 
although there is some extra scattered intensity in the mixed 
$A'_{1g}$ symmetry at high energy. Since several spots have 
been selected, this crossed procedure confirms the reliability 
of the data. Note that due to the high surface quality of the 
Hg-1223 crystal, the residual elastic scattering is confined 
below 50 $cm^{-1}, $ and is quite weak. In crossed polarization 
measurements $(B_{1g} $ and $B_{2g}), $ this scattering is just not 
seen, hence no correction has been performed to take it into 
account. The $A_{1g }+B_{2g } $  spectrum obtained from the B crystal,
is displayed in Fig.4. For the sake of the comparison, we 
show in Fig.5 preliminary raw spectra from Hg-1212. The surface 
quality being not so good for this sample, we see clearly, especially 
in the $A_{1g} $ symmetry, the Rayleigh scattering. Nevertheless, 
these results are suggestive of a very strong similarity of 
the Raman response in both compounds.

\medskip
{\it 1. The electronic  maxima}

\medskip
Let us focus on Fig. 3 and 4. Remarkably and in sharp contrast 
with Y-123 and Bi-2212, we get a large spectral range below 
the energy gap where no Raman active phonon mode hinders the 
$B_{1g } $ and $B_{2g } $  electronic scattering analysis.  Weak 
and narrow peaks appear at 240, 388, 480 and 580 $cm^{-1} $ 
in the $A_{1g }+B_{1g } $  
and $A_{1g }+B_{2g } $ but their energy location and their intensity  
are such that no correction is actually needed to discuss the 
electronic spectra. The 580 $cm^{-1} $  peak shows a phonon asymmetric 
lineshape with an antiresonance on the high energy side (see 
Fig.3-a). This is characteristic of the interference between 
an electronic continuum and a single phonon state. This Fano 
lineshape has already been detected in Y-123 and Bi-2212 for 
the $B_{1g } $ normal mode in the vicinity of the gap structure 
maximum [7,10,23]. The 640 $cm^{-1} $ peak detected in Fig.4 is 
assigned to some trace of $BaCuO_{2} $ impurity \cite{[24]}. Overall, 
the spectra do not require the delicate handling of ''phonon 
subtraction'' and we are in a position to turn immediately to 
the analysis of the data. 

Two well marked maxima are observed: one is located around 530 
$cm^{-1} $ in the $A_{1g }+B_{1g } $  (Fig.3-a), $   A_{1g }+B_{2g } $  
(Fig.4) symmetries and the $A'_{1g}$ (Fig.3-b) mixed symmetry; 
the other is observed at 800 $cm^{-1} $ in the $B_{1g} $ symmetry. 
In contrast, no clear maximum appears in the $B_{2g } $ and the 
mixed $B'_{2g}$ symmetries, for which actually the scattered 
intensity levels out smoothly around 530 $cm^{-1}. $ 

In the $A_{1g }+B_{1g } $ spectrum (Fig.3-a), we observe a weak 
shoulder near 800 $cm^{-1}, $ not seen in the mixed ${A'_{1g} } $ 
symmetry (Fig.3-b). In the $A_{1g }+B_{2g } $ spectrum (Fig.4),beyond  
the well defined maximum at 530 $cm^{-1}, $ a broad maximum is 
clearly visible at 800 $cm^{-1}. $ There is in the latter case 
no possible contamination by the $B_{1g} $ channel, whereas the 
$B_{2g} $ channel does not display any feature above 500 $cm^{-1}. $ 
All this strongly suggests that the 800 $cm^{-1} $ maximum is an 
intrinsic electronic feature in the $A_{1g} $ channel. This important 
point will be discussed later. 

The observation of two electronic maxima below $T_{c} $ is currently 
mentioned for two distinct channels [7-9, 13]. It is a common 
feature for numerous high-$T_{c} $ cuprates studied by Raman spectroscopy 
near the optimally doped regime. The lower energy peak $\omega _{l}$ 
is usually observed in $A_{1g }+B_{1g } $ symmetries and the higher energy 
one $\omega _{h}$ is detected in pure $B_{1g } $ symmetry. 
The energies of these 
two peaks for various cuprates are listed in Table I. Two well 
defined energy scales appear: (5 $-6.4)k_{B}T_{c} $ and $(8-9.4)k_{B}T_{c}.$ 
Remarkably enough, their ratio is approximatively 1.5, whatever 
the compound. What is new in our study is that, by looking at 
a less commonly studied symmetry channel, we find that these 
two energy scales are present within one single channel, namely 
$A_{1g}. $

Finally, the temperature dependence of the imaginary parts of 
the electronic response for the $A_{1g}+B_{2g} $ and $B_{1g} $ symmetry 
as a function of the temperature from 13 K to 150 K is shown 
in Fig.6-a and 6-b respectively. The imaginary part of the electronic 
response function $Im\chi [\omega ,T] $ is derived by
subtracting the photomultiplier  
dark current, the Rayleigh scattering intensity, when present, 
and by correcting for the response of the diffraction grating 
and for the Bose-Einstein factor
$n(\omega ,T)=[exp(\hbar \omega /kT)-1]^{-1}. $ In 
both symmetries, the low energy scattered intensity increases, 
approaching that of the normal state as the temperature is raised. 
The maxima at 540 $cm^{-1}$ (for $A_{1g}+ $ $B_{2g}) $ and 800 $cm^{-1} $ 
(for $B_{1g}) $ gradually collapse and disappear above $T_{c}. $ These 
two maxima do not shift significantly (at most 10\%) towards 
low energies as the temperature increases. The latter result 
is fairly general in Raman and tunneling experiments.

\medskip
{\it 2. The low lying energy excitations} 

\medskip
We wish now to concentrate on the low energy part of the spectra. 
We collect in Fig. 7 the imaginary parts of the electronic response 
functions $Im\chi [\omega ,13 K] $ obtained from the raw data after
performing the 
same corrections described just above, associated to the $A_{1g }+B_{1g }, $ 
 $   B_{1g } $ and $B_{2g } $ symmetries, already reported in \cite{[16]}. 
The dotted lines represent $Im\chi [\omega ,150 K], $ after smoothing the 
spectra (for clarity). As could be inferred from the raw spectra, 
the three response functions seem to extrapolate to zero at 
zero frequency, within our experimental accuracy. We recall 
\cite{[16]} the main observations which emerge from Fig. 7, namely: 

\medskip
i) the $B_{1g } $ response function exhibits a clear decrease 
of the electronic scattering at low energy in the superconducting 
state with respect to the normal state. In contrast, the $A_{1g }+B_{1g } $ 
and $B_{2g } $  response functions decrease barely at low energy. 
Actually, there is very little change between the normal and 
the superconducting state in the $B_{2g} $ channel. The change 
is instead much larger in the $A_{1g} $ channel but occurs around 
the peak energy.

\medskip
ii) the $B_{1g } $ spectrum exhibits a steady quasi-linear increase 
with energy, in contrast with Bi-2212 spectra where, after subtracting 
the phonon contribution, the spectrum below the peak energy 
exhibits a strong upward curvature \cite{[6]}. 
\medskip

iii) the electronic response functions in the normal and superconducting 
state merge in all three symmetries around 1000 $cm^{-1}. $
\medskip

To investigate further the low energy behavior 
of the $B_{1g} $ spectrum in the superconducting state, we have 
performed Raman measurements at T=13 K with various excitations 
lines. The imaginary parts of the electronic response functions 
in the $B_{1g } $ symmetry are displayed in Fig. 8-a for five 
different excitation lines, namely 647.1, 568.2, 514.52, 488.0 
and 476.5 nm. For clarity, the spectra have been normalized so 
as to achieve the same intensity around the 800 $cm^{-1} $ maximum 
and shifted vertically. The spectra reported here have been 
taken from sample B. There is a clear change in the overall 
shape of electronic response as the excitation energy decreases.

A simple visual inspection of the spectra suggests that they 
can be classified in two categories: for the 647.1 and 568.2 nm 
lines, the linear part appears to extend up to higher energy 
than for the 514.52, 488.0 and 476.5 nm excitation lines. Also 
the maximum for the two former ones seems to develop at a somewhat 
lower energy. In order to confirm the occurrence of two sets 
of data displaying a similar spectral dependence, we have firstly 
checked quantitatively that within the two sets, the spectra 
are indeed proportional. Then, we have used this calculation 
to normalize the spectra for the two lower energy excitation 
lines (647.1 and 568.2 nm) on one hand , and for the three higher 
energy excitation lines (514.52, 488.0 and 476.5 nm) on the other 
hand. The result is shown in Fig.8-b. Indeed, within each set, 
the higher excitation energy spectra and the lower ones fall 
exactly one on top of the other, within our experimental resolution. 

As mentioned in section II, a source of complication of the 
analysis may arise from the dependence of the Raman vertex on 
$\omega _{L} $ and $\omega _{S}. $
The dependence on $\omega _{L} $ is revealed in our 
experiments as a change in the strength of the Raman scattering 
with the laser frequency. The dependence on  $\omega _{S} $ translates 
into a dependence on the Raman shift $\omega =\omega _{L}-\omega _{S}, $ which 
could affect the validity of the low frequency analysis. 

Let us first comment on the possible origin of the ${\omega}_{L}$
dependence. It is interesting to note that in the spectral range
of interest (15000-22000 $cm^{-1}), $ various cuprates (YBCO, LaSrCuO, 
BiSrYCuO) exhibit two bands, located typically around 12000-14000 $cm^{-1} $ 
(the so-called charge-transfer band), and 22000 $cm^{-1} $ [25-28]. 
Since, to the best of our knowledge, similar spectra are not 
yet available for the Hg compounds, we assume that the absorption 
bands lie in the same range as for other cuprates. The three 
high energy excitation lines are clearly located in the absorption 
range of the high energy band. The 568.2 nm line (17600 $cm^{-1}) $ 
actually lies in-between the two bands, still closer to the 
charge transfer band, and the Raman spectrum exhibits a strong 
similarity with the 647.1 nm (15450 $cm^{-1}) $ one. The latter is 
definitely in the vicinity of the charge transfer band. Therefore, 
the change of the Raman spectra with the excitation energy bears 
an actual consistency with the absorption bands in the visible 
range, supporting the idea that resonant scatttering is effective. 
It is not very surprising that we find within our experimental 
resolution, that the data can be separated into two sets, since 
two bands seem to be involved. The question which is raised 
once this is recognized is which frequency range is most affected 
by the change of the Raman vertex with energy. When fitting 
the absorption bands to Lorentz oscillators \cite{[25]}, we found typical 
widths of 1eV. A typical energy scale for the variation of the 
Raman vertex with $\omega$ would then be 0.5-1eV.

A first order expansion of the Raman vertex (we omit the ${\bf k}$
 dependence of $\gamma _{k} $ for simplicity) with respect to the laser
 frequency 
writes: 
\begin{eqnarray}
\gamma ({\omega }_{L},\omega )=\gamma ({\omega }_{L},0)+\omega
{\left({{\partial \gamma ({\omega }_{L},\omega ) \over
\partial \omega }}\right)}_{\omega =0}
\label{eq8}
\end{eqnarray}
Therefore in the limit $ \omega \rightarrow 0 $ ,
which is of interest here, the Raman vertex is independent 
of $\omega , $ which implies that the low frequency behavior should 
not depend on the laser excitation energy. As $\omega  $ increases, 
the frequency dependence of $\gamma  $ may not be neglected. Using this 
line of reasoning, we have compared the low frequency dependence 
of the two sets of spectra obtained by varying the laser excitation. 
A simple way to do it is to select from each set the spectrum 
exhibiting the best signal to noise ratio, namely the ones associated 
to the 568.2 and the 514.5 nm lines. Fig. 9 shows that 
indeed, the low frequency dependence (up to 500 $cm^{-1}) $ is the 
same, and that the spectra gradually separate as the frequency 
increases. We assign this behavior to the change of the Raman 
vertex with frequency, as explained above. Therefore, we shall 
consider henceforward that there is no change of the low frequency 
behavior when changing the excitation energy. Note incidentally 
the strong linear term of the $B_{1g} $ spectrum, best seen in 
Fig.9.

\section{Discussion}
We may now turn to the comparison of our data with existing 
theories. We start with the $d_{x^{2}-y^{2}} $ model, as computed 
by Devereaux et al., which was found to describe well the Bi-2212 
results \cite{[22]}. In a pure $d_{x^{2}-y^{2}} $ model,
symmetry considerations 
imply that the $B_{1g } $  spectrum is insensitive to the 
nodes at $45^{0}$ (hence the $\omega ^{3} $  dependence) and displays a 
maximum at $2 {\Delta }_{0}$
, whereas the $B_{2g } $ symmetry exhibits a linear low frequency 
dependence (because it probes the nodes) and a smeared gap \cite{[16]}. 
For the same reason the $A_{1g } $ symmetry has also a linear 
low $\omega  $ behaviour. We find a fair agreement of this model with 
our data for the linear dependence of the low energy part of 
the $B_{2g } $ symmetry (see Fig. 10) . 

Our linear $A_{1g } $ spectrum agrees also quite well with the 
prediction of the $d_{x^{2}-y^{2}} $ model at low frequency, but this 
is not the case for the higher frequency range [Fig.12-a]. 
Indeed we note that the $d_{x^{2}-y^{2}} $ model has only a single maximum 
which gives rise to a single peak, both in $B_{1g } $ and $A_{1g } $ 
symmetry. Naively one expects these two peaks to be essentially 
at the same frequency. This is already in contrast with experiments 
on Bi-2212 where the $A_{1g } $ peak is found at much lower frequency 
than the $B_{1g } $ one. Nevertheless if one takes into account 
the strong screening in the $A_{1g } $ symmetry together with 
a sizeable smearing, one may obtain a large softening of the 
$A_{1g } $ peak and reconcile theory with experiment. However, 
we find it eventually difficult in our calculations to get the 
necessary softening without obtaining simultaneously an unduly 
wide maximum. We give an example of such a calculation in Fig.12-a.
Anyway our present observation of two distinct peaks in the 
same $A_{1g} $ channel deduced from the $A_{1g}+B_{2g} $ symmetry 
(see Fig. 4) cannot be reconciled with a pure $d_{x^{2}-y^{2}} $
model. We note that the second peak, essentially at the same 
frequency as the $B_{1g} $ one, had not been seen in our earlier 
experiments \cite{[16]} because only the $A_{1g}+B_{1g} $ symmetry had 
been measured.

Finally we consider the $B_{1g } $  symmetry which, together with 
the $B_{2g } $  symmetry, is easier to analyze because it is not 
affected by screening, in contrast with the $A_{1g } $ symmetry. 
Most strikingly the low frequency behavior
of the theoretical $d_{x^{2}-y^{2}} $
$B_{1g } $ spectrum is incompatible with our experimental results. 
In the $d_{x^{2}-y^{2}} $ model, the response function for $B_{1g } $ 
 should increase as $\omega ^{3} $ \cite{[6]} and it is clear, by inspection 
of Fig. 9 and 11, that our results are essentially linear in frequency. 
If we analyze our data more quantitatively, we have to recall 
that such a power law behavior is an appropriate description 
only at low frequency, and therefore the pending question is 
the choice of the proper energy scale when trying to fit the 
results. Fitting the spectrum from Fig.11 to a power law $\omega ^{\alpha }, $ 
up to 300 $cm^{-1} $ yields $\alpha =1.5\pm 0.5 $ inconsistent
with an $\omega ^{3} $  
dependence \cite{[16]}. However, the choice of the energy range for 
fitting may be disputable, and a second fit to $b\omega +c\omega ^{3} $ was 
attempted in order to provide a more accurate estimate of the 
relative weight W of the linear part with respect to the
cubic part. W was computed 
over various energy ranges, e.g. up to 300, 400, 500, 600 $cm^{-1}, $ 
still using the spectra shown in Fig.11. Actually, referring 
to Fig.9, the low energy range of interest may not exceed 500 $cm^{-1}. $ 
We found $W  = 2 \pm 1 , 4 \pm 1 , 3.3 \pm  0.7 , 3.4 \pm  0.5 $,
meaning that a dominant linear component must be taken into 
account, inconsistent with a pure $d_{x^{2}-y^{2}} $ model. $B_{1g} $ 
being linear at low frequency appears to agree with the Tl-2201 
results, at least at high enough excitation energy ( $ \hbar \omega
> $  2.18 eV, or $\lambda < $ 5734 \AA ) \cite{[28]}.

It has been argued that impurities are responsible for the observed 
low frequency density of states in the $B_{1g } $  channel. Devereaux 
\cite{[30]} has indeed shown that, in the unitary limit,
for a $d_{x^{2}-y^{2}} $
gap, impurities induce for $B_{1g } $ a linear rise of the electronic 
scattering at low frequency, crossing over to the $\omega ^{3} $  dependence 
at higher frequency. The cross-over energy $\omega ^{*} $  found in 
this case is typically the width $\gamma  $   of the band of states 
bound to impurities \cite{[31]}. For unitary scatterers this width 
is of order $\gamma  \sim (  \Gamma  \Delta _{0} $ $)^{1/2} $   
where $\Gamma  $ $= $  $n_{i} $ 
$/(\pi N_{F}) $ with $n_{i} $ the impurity concentration and  $\Delta _{0} $ 
the maximum of the d-wave gap. In the Born limit it is in general 
much smaller $\gamma \sim  4 \Delta _{0} $ $exp(-\pi  $  $\tau  $ 
$\Delta _{0}), $ where $\tau  $ is 
the normal state relaxation time, and impurities bring no qualitative 
change in the density of states at low energy\cite{[31]} except for 
large relaxation rates. Therefore the unitary case is the most 
favorable one for an important effect of impurities on Raman 
scattering. However this is a very specific case, and experimental 
support is still lacking. Note that in the case of a small or 
intermediate relaxation rate, one goes rapidly out of the unitary 
regime as soon as the scattering phase shift goes away from 
$\pi /2 $ . Only for a large relaxation rate 
$1/(2 \tau \Delta _{0}) > 0.3$ does 
the unitary regime extend over a fair range of scattering phase 
shifts. In the absence of experimental evidence, a random distribution 
of scattering phase shifts would seem a reasonable assumption.

It is worth recalling that this assumption of unitary scatterers 
was raised [32,33] in order to reconcile the robustness of the 
critical temperature against impurity scattering, which is unexpected 
in a d-wave superconductor, with the much less robust temperature 
dependence of the London penetration depth, whose linearity 
is thought to be affected by scattering. Nevertheless, due to 
the difficulty of measuring the absolute value of the penetration 
depth, there is no strong experimental evidence for this assumption. 
Actually, published data are somewhat contradictory [34,35]. 
Let us nevertheless consider this unitary limit. Our experimental 
results can be compared reasonably well with Devereaux's calculation 
\cite{[30]}, taking $\Gamma \sim 0.5 \Delta _{0}. $ This means a scattering
rate $\Gamma  $  
 of order of the gap $\Delta _{0} $ itself. Such a high scattering rate 
should strongly reduce the critical temperature or even destroy 
completely superconductivity. This is in marked contradiction 
with $T_{c}=126 K $ of our single crystals which does not differ in 
an appreciable way from the critical temperature of supposedly 
rather clean crystals. Finally we remark that the fairly sharp 
structure that we observe in $A_{1g} $ is difficult to reconcile 
with a large scattering rate. All together this makes rather 
unlikely the explanation of the linear behavior of the low frequency 
$B_{1g} $ by an impurity effect. 

Accordingly we consider that our results are representative 
of pure crystals. Then an inescapable consequence of the observed 
linear frequency dependence is the occurrence of a finite density 
of states exhibiting a similar energy dependence, which is most 
naturally assigned to nodes in the gap. Since these nodes are 
probed in both $B_{1g } $ and $B_{2g } $ symmetries, they cannot 
be located only in the $45^{0}$ direction. We are thus left with 
the conclusion that nodes exist outside the $45^{0}$ direction. It 
is worth recalling that although Raman scattering probes the 
electronic excitations along various {\bf k} directions, it 
cannot give accurately the location of the nodes \cite{[16]}. We 
have therefore to explore this point more quantitatively.

From the low energy part of our spectra, we know there are nodes 
outside the $45^{0}$ direction. It can readily be seen that a small 
shift of the nodes away from $45^{0}$ is not enough to account for 
experiment. Indeed if we use a toy model $\Delta (\theta ) = \Delta _{0} $ 
$cos(2\theta -2\alpha ) $ obtained by artificially
rotating the $d_{x ^{2}-y ^{2}} $  
order parameter by an angle $\alpha , $ we find that a
rotation by $\alpha   \approx
10^{0}$ is barely noticeable in the spectra : the linear rise at 
low $\omega  $ produced in $B_{1g } $ by this rotation is quite small
and the spectrum is essentially unchanged within experimental accuracy.

Having at this stage exhausted all 
information contained in the low energy dependence, in order to build 
up a specific model more appropriate than $d_{x^{2}-y^{2}} $, we 
have to rely now on the high energy part of our spectra. While 
the $B_{2g} $ symmetry offers only a hint of a maximum around
$500 \, cm^{-1}, $ 
the $B_{1g} $ symmetry and the $A_{1g} $ symmetry are much more informative.  
Indeed we have the occurrence of two maxima at 500 and $800 \, cm^{-1}, $ 
seen simultaneously in pure $A_{1g} $ symmetry. The last one is 
seen separately in $B_{1g} $ in a very clear way while the former 
is compatible with the structure seen in the $B_{2g} $ symmetry. 
Together these observations strongly suggest two different energy 
amplitudes for the gap.

We assume that the gap has the symmetry of one of the one-dimensional 
representation of the tetragonal group, namely $A_{1g}, $ $B_{1g}, $ 
$B_{2g} $ or $A_{2g}. $ We need only to consider the range of 
the Fermi surface where the wavevector {\bf k} has an angle 
$\theta  $ with the x axis between 0 and $\pi /4. $ The rest can be deduced 
by symmetry operations. Our model must accommodate a main maximum 
M, a secondary maximum M' - for the absolute value of the order 
parameter - and at least one node N (we assume for simplicity 
that there is only a single secondary maximum and no secondary minimum
since there is no experimental evidence for them). This does not match with 
any of the simplest representative for the order parameter, 
namely $k_{x}^{2}+k_{y}^{2} $ $(A_{1g} $ symmetry ),
$(k_{x}^{2}-k_{y}^{2}) $  
$(B_{1g} $ symmetry ), $k_{x } $ $k_{y } $ $(B_{2g} $
symmetry ), or $k_{x } $  
$k_{y } $ $(k_{x}^{2}-k_{y}^{2}) $ $(A_{2g} $ symmetry ). For 
instance, as pointed out by Leggett, the simple $A_{2g} $ symmetry 
is compatible with the low energy part of our spectra; but it 
does not provide a proper account of the high energy part since 
it would give only peaks at a single frequency corresponding 
to the maximum of the gap. Therefore we have to deal with a 
more complicated order parameter. From symmetry we must find 
at $0^{0}$ either M or M' or N, and similarly at $45^{0}$ ( we eliminate 
the possibility of a discontinuous order parameter ). Since 
the main maximum M is seen in $B_{1g} $ symmetry and not in $B_{2g} $ 
symmetry, we have to put it at $0^{0}$ . Similarly the secondary 
maximum M' being inferred in $B_{2g} $ symmetry but not seen 
in $B_{1g} $ symmetry 
implies a location at $45^{0}$. Then the only place left for nodes 
is somewhere between $0^{0}$ and $45^{0}$. Since we have assumed a single 
secondary maximum, we can put only one node there. This leads 
to an order parameter which has over the whole Fermi surface 
4 (equivalent) M, 4 M' and 8 nodes. Since this order parameter 
is continuous and is nonzero at $0^{0}$ and $45^{0}$, it has the $A_{1g} $ 
symmetry. Therefore we come to the conclusion that our Raman 
scattering data imply an order parameter which does not break 
spontaneously the crystal tetragonal symmetry. 
However it is clear that any slight shift of M and M' away from 
their symmetric locations will essentially not be felt in the 
resulting spectra. For example since our $B_{2g} $ is quite featureless, 
we may worry that the presence of the main maximum M is washed 
out in it. So M could be away from $0^{0}$. Since M' is at $45^{0}$, we 
can only put a node at $0^{0}$. But there is a node away from $0^{0}$ 
because $B_{2g} $ is linear at low frequency. The only location 
for this node is between M and $45^{0}$. We end up with an order 
parameter which has the $B_{2g} $ $(d_{xy}) $ symmetry, with 8 M, 
4 M' and 12 nodes over the whole Fermi surface. Similarly we 
might worry that a small contribution of M' to the $B_{1g} $ spectrum 
is not seen in our experimental results. However this seems 
almost incompatible with our data, except if M' is quite near 
$45^{0}$, but this would lead to an almost discontinuous order parameter 
since we have to put a node at $45^{0}$. In this way we would have 
a $B_{1g} $ symmetry order parameter, with 4 M, 8 M' and 12 nodes 
on the Fermi surface. Finally we could also think of putting 
nodes at $0^{0}$ and $45^{0}$ , with M and M' slightly away from these 
locations. We would need to put a node between them. Hence we would 
have a $A_{2g} $ order parameter, with 8 M, 8 M' and 16 nodes. 
Again this is hardly compatible with our $B_{1g} $ data. In conclusion 
the $A_{1g} $ order parameter is at the same time the simplest 
and the most compatible with our data. In the following we will 
concentrate on this symmetry and investigate it more quantitatively.

A simple one-parameter model having the $A_{1g} $ symmetry and 
satisfying the above requirements is $\Delta (\theta )
 =   \Delta _{0} $  $[\cos $  
$(4\theta )+s] $ where s = ( $\Delta _{M} $  - $\Delta _{m} $  )
/ $(\Delta _{M  +} \Delta _{m }) $ 
and  $\Delta _{0} $ = ( $\Delta _{M} $  + $\Delta _{m } $  ) / 2 .
The gap maximum  
$\Delta _{M} $   is obtained for $\theta  $  = 0 while the secondary maximum 
$   \Delta _{m } $ occurs for $\theta  $    = $\pi /4 $ .
The node lies at $\theta _{0}=(1/4) $  
arccos(-s) and in this simple model its location is linked to 
the ratio $\Delta _{m } $  $/\Delta _{M} $   . Naturally by introducing other  
parameters we can decouple these two quantities, as we will 
consider briefly below. Since we make the choice s = 0.27 according 
to experiment, we find obviously the peaks for $B_{1g}$ and $B_{2g}$ in 
agreement with our experimental results (Fig. 10 and 11).
We calculate the Raman 
spectra for $B_{1g}$ and $B_{2g}$ symmetry by taking for the Raman vertices 
the simplest form compatible with symmetry, namely $\gamma _{B_{1g} }
 \sim \cos 2\theta  $ , $\gamma _{B_{2g} } \sim \sin 2\theta  $
and a density of states independent 
of $\theta . $ Naturally the bare result for these spectra display a 
logarithmic singularity at their respective maximum and we have 
to invoke some broadening, most likely due to lifetime effects, 
in order to obtain theoretical results compatible with experiment. 
In this paper we include this broadening at the most phenomenological 
level by merely convoluting the bare result with a broadening 
function. We take a simple normalized lorentzian for this function. 
However we expect lifetime effects to be quite small at low 
energy and to grow progressively at higher energy. We account 
for this by choosing the width $\Gamma (\omega ) $
of our lorentzian proportional  
to the frequency at which we are looking :
$\Gamma (\omega ) $ = a $\omega  $  .
Explicitly our final spectrum $F(\omega ) $ is obtained from the bare 
spectrum $B(\omega ) $ by $F(\omega ) $ =
$\int $ d$\omega ' $ $L(\omega  $  $- $  $\omega ', $
 $\omega ) $ $B(\omega ') $ 
where $L(\omega  $  $- $  $\omega ', $  $\omega ) $ =
(1/  $\pi ) $ $\Gamma (\omega ) $
/((  $\omega  $  $- $  $\omega ')^{2}+ $ 
 $\Gamma ^{2}(\omega )) $ . Here the integration goes from $ - \infty $ to 
$+ \infty $ and $B(\omega ) $ is extended by
symmetry $B(-\omega ) $ = $- $ $B(\omega ) $  
in the whole frequency range. This broadening method insures 
F(0) = 0 , whatever $\Gamma (\omega ). $ In Fig.10 and 11  we have taken our 
broadening parameter a = 0.15 . Note that this parameter is 
our only way to adjust the peak height with respect to the low 
energy part of the spectrum. Hence the fact that we find a good 
agreement in $B_{1g} $ with experiment with a reasonably small 
broadening is satisfactory (Fig.11). This is directly due to 
the fact that our node location $\theta _{0} $  = $26^{0}$ is far enough 
from $45^{0}$, otherwise we would have trouble obtaining a strong 
enough low energy spectrum for $B_{1g}, $ as we have seen in our 
toy model. As expected, the part of the experimental spectrum 
above the peak frequency is not properly described by our model, 
inasmuch as we go toward higher energy. This is merely due to 
the fact that our theoretical result goes to zero for very high 
energies, in contrast with experiment. This discrepancy would 
be cured, at least partially, by taking proper account of lifetime 
effects, but this is beyond the scope of the present paper. 
In particular we ascribe the moderate agreement between theory 
and experiment in $B_{2g} $ symmetry to strong lifetime effects 
in this channel (Fig.10).

We consider now in detail the $A_{1g} $ symmetry. In our theoretical 
analysis, we take screening fully into account in the standard 
way. This feature makes the interpretation of the $A_{1g} $ spectrum 
less easy than the $B_{1g} $ or $B_{2g} $ spectrum, because the result 
is not systematically related in an intuitive way to the shape 
$\Delta (\theta ) $  of the order parameter. The most obvious manifestation 
of this is the fact that a constant $A_{1g} $ Raman vertex gives 
a zero contribution because of full screening. Therefore we 
have to look for the next term in a Fourier expansion in order 
to obtain a nonzero result. This term is proportional to $\cos 4\theta . $ 
However it goes to zero for $\theta  $  $=\pi /8, $ that is very near the 
location of the node in our specific model. This implies that 
the $A_{1g} $ electronic response will be quite small at small 
energy, in contradiction with experiment, which shows a strong 
linear dependence. For a sizeable $A_{1g} $ response at low energy, 
we need the zero of the $A_{1g} $ Raman vertex and the node of 
the gap to be clearly apart. Since we have already chosen our 
order parameter, hence the node location, we achieve this by 
considering a more complicated Raman vertex, e.g. including 
the next order Fourier component : we take
$\gamma _{A_{1g} } \sim A \cos 4\theta -(1-A) \cos 8\theta . $  
As seen in Fig.12-b, we obtain a reasonably good agreement with 
experiment for A = 0.05 and a small smearing a = 0.03 . We note 
that we are led to use different smearing for the different 
Raman symmetries, in contrast to what has been done in \cite{[16]}. 
This does not make a problem: the smearing is meant to account 
for lifetime effects at a phenomenological level. Hence when 
dealing with different channels, there is no reason to have 
the same effective scattering. We obtain a sizeable contribution 
at low energy, at   $\Delta _{m } $ and at $\Delta _{M} $  : this is clearly 
related to the fact that the Raman vertex is not small for $\theta  $ 
 =0 , for $\theta  $  $=\pi /4 $ and at the node $\theta  $
$=\theta  $  $_{0}. $ Note that, 
while the size of the low frequency part is directly related 
to the strength of the Raman vertex around the node, the same 
is not true for the strength of the peaks at  
$\Delta _{m } $ and $\Delta _{M}. $  
In particular one may obtain a single broad bump between $\Delta _{m } $ 
and $\Delta _{M} $  instead of two peaks. This is naturally due to 
the complex effect of the screening term.

The small value A=0.05, which implies that the $\cos 4\theta  $ contribution 
is negligible in the Raman vertex compared to the
$\cos 8\theta  $  contribution,  
comes from the strong constraints that the various characteristics 
of our $A_{1g}$ spectrum put on the order parameter. We had already 
been led to consider a similar vertex in ref.\cite{[16]} to account 
for a $A_{1g} $ spectrum which did not display a maximum at $\Delta _{M} $.
While this is not expected beforehand, this does not imply 
anything pathological about the Raman vertex we use and such 
a situation may very well occur accidentally. As we mentioned 
we are merely led to this case by our request of a sizeable 
vertex at 0 , $\pi  $ /4 and at the node. Clearly we can achieve 
this goal in other ways if we allow more flexibility in our 
order parameter. 

We demonstrate this by considering the following order parameter 
: $\Delta (\theta ) $  $= $  $\Delta _{M} $  
for 0 $< $ $\theta  $  $< $ $\theta _{0}- \Delta _{M} $ /S 
,  $\Delta (\theta ) $  $= $ S
( $\theta _{0}  - \theta  $ ) for $\theta _{0}- \Delta _{M} $ /S $< $ 
$\theta  $  $< $ $\theta _{0} $ + $\Delta _{m } $ /S
, $\Delta (\theta ) $  $= $
 $- $  $\Delta _{m } $   for 
$\theta _{0} $ + $\Delta _{m } $ /S $< $ $\theta  $  $< $ $\pi /4 $ .
Its shape is similar to 
the simple one that we have used above, but in this way we can 
choose more freely the location of the node $\theta _{0} $ 
of $\Delta (\theta ) $  
and its slope S. On the other hand we restrict ourselves to 
the simplest Raman vertex  $\gamma _{A_{1g} } \sim \cos 4\theta  $ . We find a 
reasonably good agreement with experiment ( see Fig.12-b ) for 
$\theta _{0} $  = $35^{0}$ and S = 3.6. This shows that we can account for 
the $A_{1g} $ spectrum by the simplest Raman vertex together with an order 
parameter having two maxima and a node. However the bare $B_{1g} $ 
spectrum provided by this latter order parameter has a very 
strong peak at $\Delta _{M} $ (inasmuch as we have taken a simple flat 
maximum model ) and a quite small low energy part, due to the 
node proximity of $45^{0}$. This can not be reconciled with experiment 
by any reasonable smearing. Again the strength of the linear 
low energy part of the $B_{1g} $ spectrum shows that the node is rather 
far away from $45^{0}$. Note however that we could cure somewhat 
this difficulty by assuming a more complicated $B_{1g} $ Raman 
vertex, which would have a small strength for $\theta  $  = 0 ( in addition 
of being zero by symmetry for $\theta  $  = $\pi /4 $ ). In conclusion we 
believe that it is possible to account for experiment for all 
symmetries by an order parameter having the qualitative features 
that we have considered, together with Raman vertices reasonably 
close to the simplest one. This might require perhaps that we 
use a varying density of states, while here we have assumed 
it to be constant for simplicity. However we have not tried 
to explore this problem more quantitatively.

Our model gives a natural explanation of the different positions 
of the $B_{1g } $ and $B_{2g } $ peaks and the existence of these 
two peaks in $A_{1g} $ symmetry (even if screening complicates the matter). 
Note that peaks with different positions are also found in the 
other high-$T_{c} $ superconductors (see table I). At this point, 
we can notice that the existence of two electronic structures 
is not only detected in Raman scattering (see table I) but also in
Giaever tunneling \cite{[36]} and spectroscopy tunneling measurements 
\cite{[37]} in YBCO. The energies of these two structures were found 
at $\Delta =20meV $ and $\Delta =30 meV $ which corresponds
to $2\Delta =320 $ and 480 $cm^{-1}$ 
respectively (very close to the values reported for YBCO in 
table I). A d+s  wave model as developed by M.T.Beal-Monod et 
al. \cite{[38]} accounts for the $B_{2g} $ spectrum but cannot explain 
the $A_{1g} $ and $B_{1g} $ spectra unless one introduces a very large
scattering rate, which is incompatible with our critical temperature
as we have discussed above. We note finally that the high 
values of the maxima ($ \sim 8-9 k_{B} T_c$) as well as the non observation
of a
significant softening of the maxima, as a function of the temperature,
as reported in section V.1, are incompatible with a weak coupling theory 
\cite{[39]}.

\section{Conclusion}
In conclusion, we have presented pure electronic Raman spectra 
of Hg-1223 single crystals. In the superconducting state our 
most significant results are i) the two electronic maxima usually 
observed separately at 530 and 800 $cm^{-1} $ in $B_{2g} $ and $B_{1g } $ 
symmetries respectively, are seen simultaneously in $A_{1g} $ symmetry. 
The existence of such two electronic features  at $2\Delta /k_{B}Tc= $ (5-6) 
and (8-9), are reported not only in Raman but also on tunneling 
measurements. ii) these two electronic structures disappear 
at $T_{c} $ without showing significant softening as the temperature 
is raised. iii) the electronic response function exhibits an 
intrinsic linear $\omega  $ dependence - not only in the $B_{2g } $ spectrum 
- but also in the $B_{1g } $ spectrum, which has been extensively 
studied in this work: the linear dependence is robust against 
the change of the excitation frequency, implying that this is 
an intrinsic feature arising from the gap symmetry. These experimental 
results strongly suggest a very anisotropic superconducting 
gap, involving two characteristic energies and nodes located 
away from the [110] and [1$\bar{1}$0] directions. Such observations 
are inconsistent with a pure $d_{x^{2}-y^{2}} $ order parameter symmetry.
The simplest order parameter compatible with our experimental data displays 
two maximum gap values and 8 nodes. It has the $A_{1g} $ symmetry. 
We have developed a detailed comparison with the experimental 
spectra and we have shown that such an order parameter can account 
satisfactorily for our experimental results.

\section{Acknowledgements}
We thank P. Monod, M. Cyrot and G. Deutscher for very fruitful 
discussions.

$\bf {Figure Captions}$ \\

FIG.1  DC magnetization recorded during field cooling in a 
10 Oe field for A and B samples \\

FIG.2  Experimental configurations chosen to obtain the $B_{1g} $, 
$B_{2g} $ and $B'_{2g} $ symmetries \\

FIG.3  Raw Raman spectra of the Hg-1223 (sample A) at 
T= 13 K obtained from both configuration described in Fig.2. 
a)$A_{1g}+B_{1g}, $ pure $B_{2g} $ and $B_{1g} $ symmetries (obtained 
by rotating the crystal).  
b) mixed ${A'_{1g}}$  and ${B'_{2g}}$ symmetries (obtained by rotating 
the polarization) \\

FIG.4  Raw Raman spectrum of Hg-1223 (sample B) at T= 13 K 
in $A_{1g}+B_{2g} $ symmetry. Note the shoulder which develops 
on the high energy side, at 800 $cm^{-1}. $ \\

FIG.5  Raw Raman spectra of Hg-1212 taken at 13K (shown for 
comparison with Fig.3) \\

FIG.6  Imaginary part of the response functions at various temperatures 
(sample B)    a)$A_{1g}+B_{2g} $ spectrum    b)$B_{1g} $ spectrum \\

FIG.7  Imaginary parts of the response function 
$Im[\chi (\omega ,T=13K)] $  
deduced from experimental data of Fig. 3. The dashed lines represent
$Im[\chi (\omega ,T=150K)]$. \\

FIG.8  Imaginary part of the response function $Im[\chi (\omega ,T=13K)] $
in $B_{1g} $ symmetry, obtained from various excitation laser lines
(sample B). Zero amplitude is indicated by the short solid line.
a)after normalization of the amplitude at the peak energy.
b)a different normalization (see text) shows that the data
can be gathered into two sets of spectra displaying the same
frequency dependence \\

FIG.9 Imaginary part of the electronic response for two excitation 
lines (514.5 nm, circles, and 568.2 nm, triangles) previously 
shown in Fig.8-b as belonging to the two sets of data. Within 
a scale factor, it appears clearly in this plot that the low 
frequency response exhibits the same frequency dependence up 
to 500 $cm^{-1}, $ for both excitation lines \\

FIG.10  $B_{2g} $ data (sample A) 
and calculations for the $d_{x^{2}-y^{2}} $ model (dashed line) and 
for a model with 8 nodes lying at $\theta _{0}$ =$26^{0}$ (solid line) \\

FIG.11  $B_{1g} $ data (sample A) and calculation for the
$d_{x^{2}-y^{2}} $ model (dashed line) and for a model with 8 nodes lying
at $\theta_{0} = 26^{0}$ (solid line). \\

FIG.12  $A_{1g}+B_{2g} $ spectrum (sample B) and calculations\\ 
a) for the $d_{x^{2}-y^{2}} $ model (solid line)\\
b) s=0.27  A=0.05  a=0.03 (solid line) S=3.6
 $\theta_{0} $ = $35^{0}$ a=0.05 (dashed line)  \\

\bigskip

\begin{tabular}{llllll}  \hline
 & & & & & \\
 & Y-123 [7] \hspace{.2in} & Bi-2212 [9] \hspace{.2in} &
LaSr(x)-214 [12] \hspace{.2in} & Tl-2201 [11] \hspace{.2in} &
Hg-1223 \\  \hline
 & & & & & \\
 Stoechiometry \hspace{.2in} & 6.95 & (see ref.) & x=0.17 & (see ref.) &
see section III \\  \hline
 & & & & & \\
 $T_{c}$ (K) & 90 & 86 & 37 & 80 & 127 \\  \hline
 & & & & & \\
 ${\omega}_{l}$ ($cm^{-1}$) & 340 & 350 & 125 & 300 & 540 \\  \hline
 & & & & & \\
 ${\omega}_{h}$ ($cm^{-1}$) & 550 & 500 & 200 & 450 & 800 \\  \hline
 & & & & & \\
 $\hbar {\omega}_{l} / k_{B} T_{c}$ & 5.6 & 6.1 & 5.1 & 5.6 & 6.4  \\ \hline
 & & & & & \\
 $\hbar {\omega}_{h} / k_{B} T_{c}$ & 9.1 & 8.7 & 8.1 & 8.4 & 9.4 \\  \hline
 & & & & & \\
${\omega}_{h} / {\omega}_{l}$ & 1.6 & 1.4 & 1.6 & 1.5 & 1.5 \\    \hline
\end{tabular}

\begin{table}
\caption{Overview of the energy location of the two electronic
maxima observed by Raman spectroscopy in various cuprates near
the optimally doped regime.}
\label{Tab1}
\end{table}

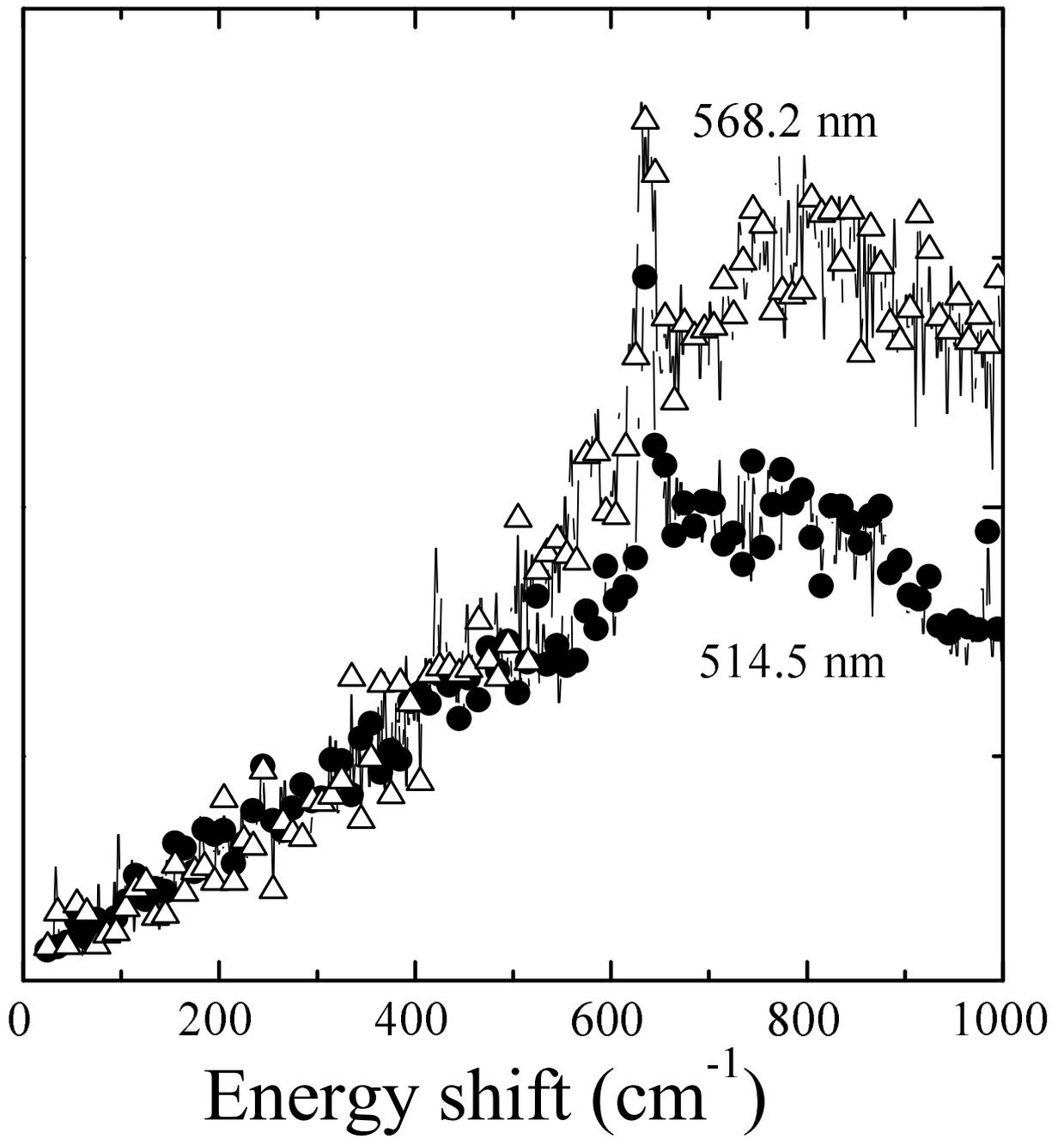
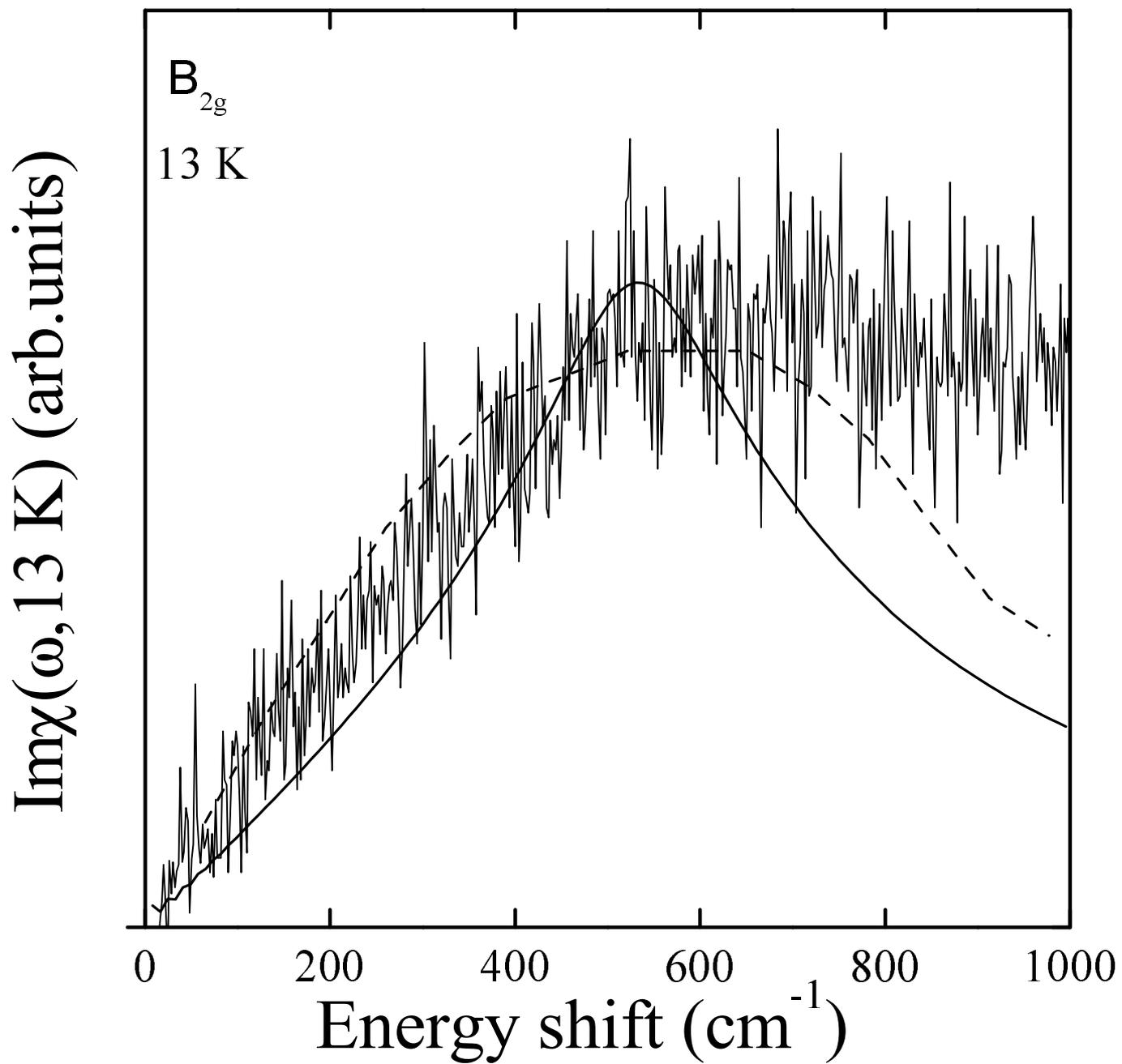
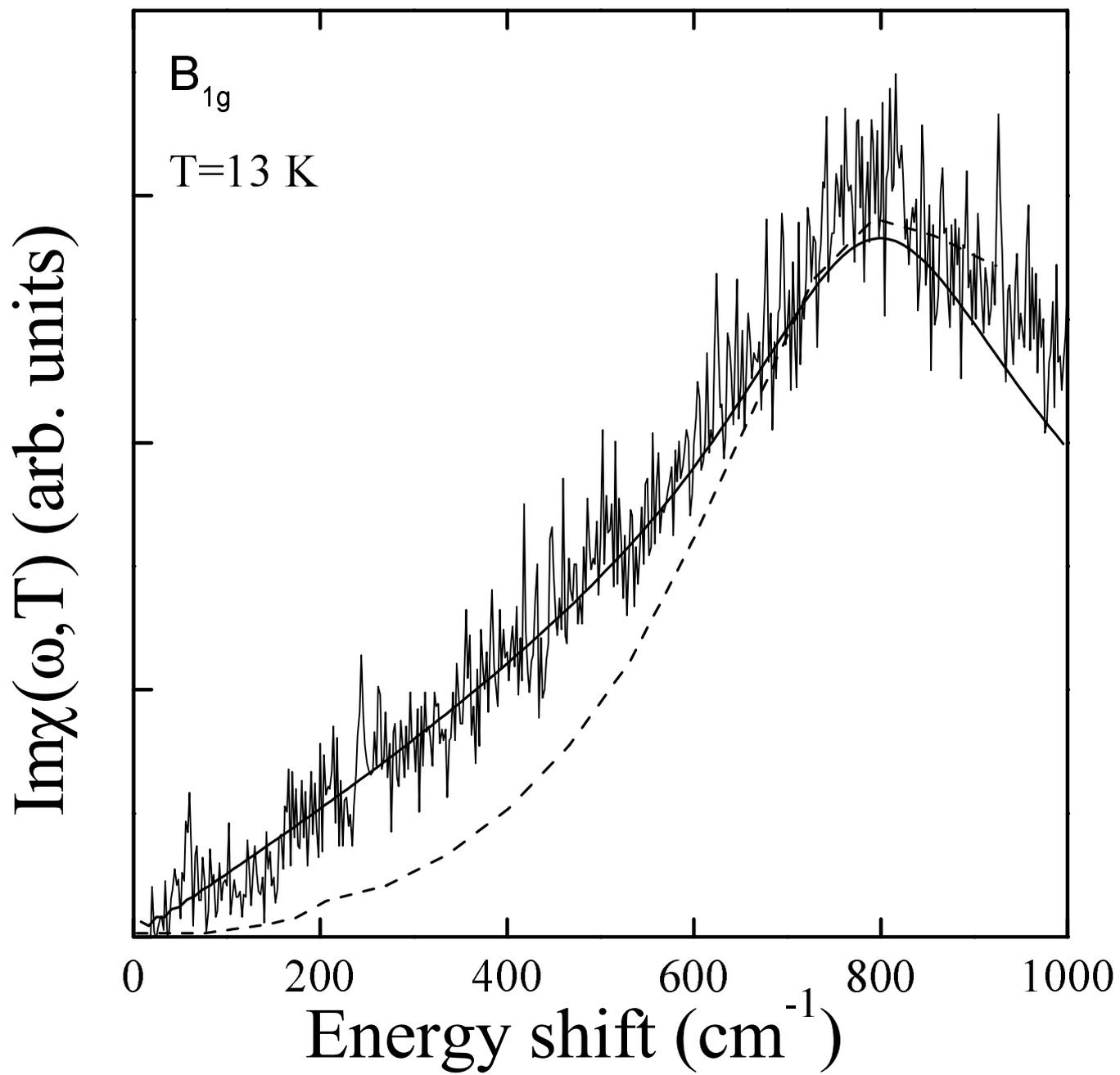
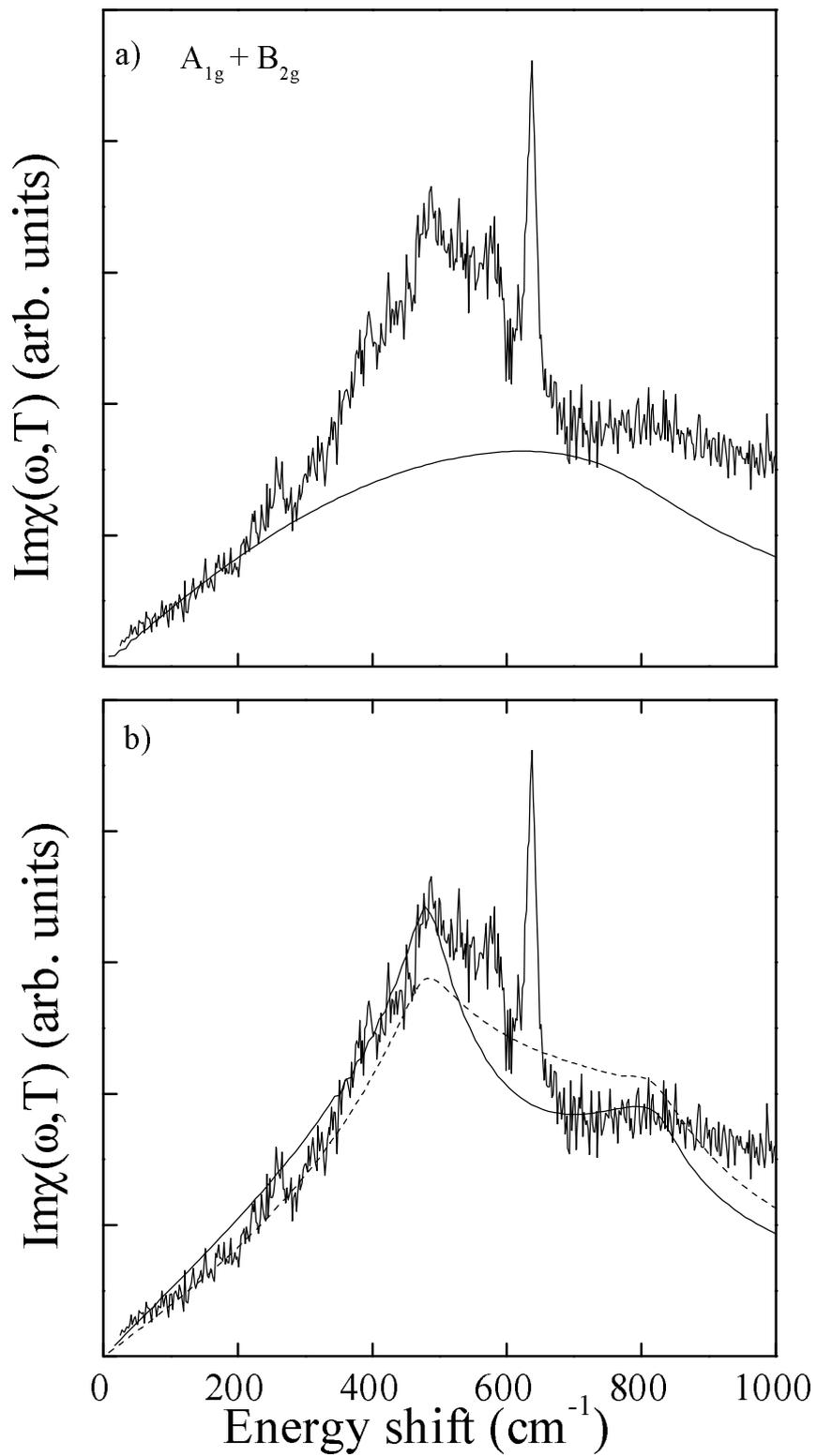
\end{document}